\documentclass[aps,prl,twocolumn]{revtex4-2}

\usepackage{amsmath,amssymb,bm}
\usepackage{graphicx}
\usepackage{physics}
\usepackage{hyperref}

\DeclareUnicodeCharacter{00AD}{}
\begin{document}
\title{Helical Current of Propagating Dirac Electron and Geometric Coupling to Chiral Environments}

\author{Ju Gao}
\email{jugao@illinois.edu}
\author{Fang Shen}%
\affiliation{Department of Electrical and Computer Engineering, University of Illinois at Urbana--Champaign, Urbana, Illinois 61801, USA}

\date{\today}

\begin{abstract}
We show that a propagating Dirac electron with intrinsic spin generically carries a real--space helical conserved current, even in the absence of orbital angular momentum.
Using exact Dirac eigenstates in cylindrical confinement, we demonstrate that this helical structure possesses definite handedness, persists into evanescent regions, and is characterized by a geometric helix pitch independent of the longitudinal de~Broglie wavelength. This intrinsic helical geometry enables a local geometric coupling between a propagating electron and a chiral environment, yielding chirality--dependent spin selectivity through current geometry rather than through a spin--orbit coupling term.
\end{abstract}

\maketitle

\section{Introduction}

Helical electronic structures are no longer purely theoretical. Recent advances in
ultrafast electron microscopy and electron--light interactions have enabled direct
visualization of spatially helical electron currents and measurement of their
longitudinal pitches~\cite{Grillo2018,Clark2013}. In most existing realizations,
however, such helicity is externally engineered and typically spin independent,
arising from structured electromagnetic fields or imposed orbital angular momentum.

At the same time, spin--selective transport and helical current patterns have been
observed in chiral and molecular systems that do not exhibit strong spin--orbit
coupling~\cite{NaamanWaldeck2015,Yang2021}. These observations suggest that intrinsic
spin may play a more direct spatial role in electronic transport than is commonly
assumed, independent of externally imposed structure or orbital motion. Despite
substantial experimental progress, the microscopic origin of chirality--induced
spin selectivity (CISS) remains actively debated and strongly model dependent,
motivating the exploration of alternative mechanisms grounded in real--space
current structure~\cite{Foo2025MindTheGapCISS}.

In free space, the spatial structure associated with the conserved current of a
propagating Dirac electron disperses rapidly. Confinement therefore provides a
natural setting in which intrinsic spin--dependent current geometry can be
stabilized and examined. In this work, we analyze the conserved Dirac current of an
electron propagating in a cylindrical confinement and show that intrinsic spin
alone generates a real--space helical current even when the orbital angular
momentum vanishes (\(l=0\)).

The resulting helix is neither a particle trajectory nor a phase artifact, but a
geometric property of the electron wave itself with direct electromagnetic
significance. This intrinsic helical current structure enables a local coupling
between a propagating electron and a chiral environment, providing a concrete
microscopic basis for chirality--dependent spin selectivity.

In this context, "without invoking spin--orbit coupling" means that no explicit spin--orbit interaction term is introduced in the Hamiltonian, and no spin precession is assumed to be induced by the chiral medium. Instead, the chirality dependence arises from the intrinsic geometry of the conserved Dirac current,
which already encodes spin as part of the relativistic current itself. Complementary to spin--orbit--based descriptions, our work identifies a geometric account of chiral spin selectivity.

\section{Dirac Eigenstates in Cylindrical Confinement}

We consider the time-dependent Dirac equation in a cylindrically symmetric confinement,
\begin{equation}
i\hbar \frac{\partial}{\partial t}\Psi(\bm r,t)
=\left[-i\hbar c\,\bm{\alpha}\!\cdot\!\bm{\nabla}
+\gamma^{0}mc^{2}+U(\rho)\right]\Psi(\bm r,t),
\end{equation}
with
\begin{equation}
U(\rho)=
\begin{cases}
0, & 0<\rho<R,\; -d<z<d,\\
U, & \rho>R,\; -d<z<d,
\end{cases}
\end{equation}
and translational invariance along \(z\).
Exact propagating solutions exist both inside the confinement and in the evanescent exterior region.

For the lowest radial mode with zero orbital angular momentum (\(l=0\)), the propagating spin-up and spin-down eigenstates take the form
\begin{equation}
\Psi_{\uparrow}(\rho,\phi,z)=
\begin{cases}
N
\begin{pmatrix}
J_0(\zeta\rho)\\
0\\
\eta k J_0(\zeta\rho)\\
i\eta \zeta e^{i\phi} J_1(\zeta\rho)
\end{pmatrix}
e^{ikz}, & \rho \le R,\\[8pt]
\kappa N
\begin{pmatrix}
K_0(\xi\rho)\\
0\\
\eta k K_0(\xi\rho)\\
i\eta \xi e^{i\phi} K_1(\xi\rho)
\end{pmatrix}
e^{ikz}, & \rho>R,
\end{cases}
\end{equation}
and
\begin{equation}
\Psi_{\downarrow}(\rho,\phi,z)=
\begin{cases}
N
\begin{pmatrix}
0\\
J_0(\zeta\rho)\\
i\eta \zeta e^{-i\phi} J_1(\zeta\rho)\\
-\eta k J_0(\zeta\rho)
\end{pmatrix}
e^{ikz}, & \rho \le R,\\[8pt]
\kappa N
\begin{pmatrix}
0\\
K_0(\xi\rho)\\
i\eta \xi e^{-i\phi} K_1(\xi\rho)\\
-\eta k K_0(\xi\rho)
\end{pmatrix}
e^{ikz}, & \rho>R,
\end{cases}
\end{equation}
where \(N\) is a normalization constant, \(\kappa\) is the interior--exterior matching coefficient, and
\begin{equation}
\eta=\frac{\hbar c}{\mathcal{E}+mc^{2}} .
\end{equation}

The radial wave vectors in the respective regions are
\begin{align}
\zeta^{2}&=\frac{\mathcal{E}^{2}-m^{2}c^{4}-\hbar^{2}c^{2}k^{2}}{\hbar^{2}c^{2}},\\
\xi^{2}&=-\frac{(\mathcal{E}-U)^{2}-m^{2}c^{4}-\hbar^{2}c^{2}k^{2}}{\hbar^{2}c^{2}},
\end{align}
and the eigenenergy \(\mathcal{E}\) is fixed by boundary matching at \(\rho=R\),
\begin{equation}\label{eq:boundary}
J_0(\zeta R)=\kappa K_0(\xi R), \quad
\zeta J_1(\zeta R)=\kappa \xi K_1(\xi R).
\end{equation}

The local charge and current densities follow from the conserved Dirac four-current
\(j^\mu=-ec\,\bar{\Psi}\gamma^\mu\Psi\)~\cite{Dirac1928,BjorkenDrell1964}.
For both spin polarizations the charge density is independent of \(\phi\) and \(z\),
\begin{align}
&q(\rho)=\nonumber \\
&-eN^{2}
\begin{cases}
J_0^{2}(\zeta\rho)\!\left(1+\eta^{2}k^{2}\right)+\eta^{2}\zeta^{2}J_1^{2}(\zeta\rho),
& \rho \le R,\\
\kappa^{2}\!\left[
K_0^{2}(\xi\rho)\!\left(1+\eta^{2}k^{2}\right)+\eta^{2}\xi^{2}K_1^{2}(\xi\rho)
\right],
& \rho>R.
\end{cases}
\end{align}

For both spin polarizations the radial current vanishes identically, \(j_\rho=0\), while the longitudinal current is
\begin{equation}
j_z(\rho)=
\begin{cases}
-2ec\,N^{2}\eta k\,J_0^{2}(\zeta\rho), & \rho \le R,\\
-2ec\,N^{2}\kappa^{2}\eta k\,K_0^{2}(\xi\rho), & \rho>R.
\end{cases}
\end{equation}
The azimuthal current for the spin-up state is
\begin{equation}\label{eq:j_phi}
j_{\phi}(\rho)=
\begin{cases}
-2ec\,N^{2}\eta\zeta\,J_0(\zeta\rho)J_1(\zeta\rho), & \rho \le R,\\
-2ec\,N^{2}\kappa^{2}\eta\xi\,K_0(\xi\rho)K_1(\xi\rho), & \rho>R,
\end{cases}
\end{equation}
while the spin-down state carries the opposite azimuthal current, \(-j_{\phi}(\rho)\).

These exact eigenstates and current components provide the complete local input for the following analysis of intrinsic helical current geometry at \(l=0\).

\section{Intrinsic Helical Current at $l=0$}

While the charge density associated with a propagating Dirac electron remains non-helical, the conserved Dirac current exhibits a spatially helical structure even at zero orbital angular momentum. This contrasts with conventional intuition, which attributes spatial helicity to orbital motion, whereas here it arises as a direct consequence of intrinsic spin in longitudinal motion.

The geometry of the current flow is characterized by its streamlines, defined locally by
\( d\bm r \parallel \bm j \).
The resulting helical streamlines describe a spatial organization of the conserved current, rather than a particle trajectory.
In cylindrical coordinates this condition yields
\begin{equation}
\frac{d\phi}{dz}=\frac{j_\phi(\rho)}{\rho\,j_z(\rho)} .
\end{equation}
For the spin-up state described by Eq.~\eqref{eq:j_phi}, this relation integrates to
\begin{equation}
\phi(\rho;z)=\phi(0)+
\begin{cases}
\dfrac{\zeta\,J_0(\zeta\rho)J_1(\zeta\rho)}
{\rho\,k\,J_0^2(\zeta\rho)}\,z, & \rho<R,\\[1.2ex]
\dfrac{\xi\,K_0(\xi\rho)K_1(\xi\rho)}
{\rho\,k\,K_0^2(\xi\rho)}\,z, & \rho>R,
\end{cases}
\label{eq:helix_solution}
\end{equation}
where \(\phi(0)\) is the azimuthal angle at \(z=0\).
The helical geometry therefore follows directly from the local current components once spin and longitudinal motion coexist. The spin-down state exhibits the opposite helical handedness. 

Both the azimuthal and longitudinal current components persist into the evanescent region \(\rho>R\), and the corresponding streamlines retain the same helical geometry as within the confinement~\cite{GaoShen2024_Entropy}. The intrinsic helicity of the conserved current is therefore not restricted to the classically allowed region, but reflects a persistent spatial structure of the Dirac current.

To identify the dominant contribution to the helical current flow, we determine the radius at which the azimuthal current density is maximal,
\begin{equation}
\frac{d\,j_\phi(\rho)}{d\rho}
=\frac{d}{d\rho}\!\left[J_0(\zeta\rho)J_1(\zeta\rho)\right]=0 .
\end{equation}
This condition reduces to
\begin{equation}
J_0(\zeta\rho_*)\!\left[J_0(\zeta\rho_*)-J_2(\zeta\rho_*)\right]
=2J_1^2(\zeta\rho_*) ,
\label{eq:rhostar}
\end{equation}
which defines a characteristic radius \(\rho_*\) serving as a natural geometric reference for the intrinsic helical current structure. This radius identifies where the intrinsic helical structure is most strongly expressed, rather than introducing a new length scale.

As a representative parameter set, we consider a nanotube of radius \(R=1\,\mathrm{nm}\), barrier height \(U=2\,\mathrm{eV}\), and electron kinetic energy \(E_k=\hbar^2 k^2/(2m)=25\,\mathrm{meV}\), corresponding to a longitudinal de~Broglie wavelength of \(\lambda_z \simeq 7.76\,\mathrm{nm}\). Using the boundary condition in Eq.~\eqref{eq:boundary}, we obtain a transverse mode energy \(\mathcal{E}_\perp=\mathcal{E}-mc^2-E_k=170\,\mathrm{meV}\) for the ground state. Solving Eq.~\eqref{eq:rhostar} yields \(\rho_*=0.51\,\mathrm{nm}\), confirming that the maximal azimuthal current resides within the confinement.

The streamline passing through \(\rho=\rho_*\) is therefore
\begin{equation}
\phi(\rho_*;z)=
\dfrac{\zeta\,J_0(\zeta\rho_*)J_1(\zeta\rho_*)}
{\rho_*\,k\,J_0^2(\zeta\rho_*)}\,z ,
\label{eq:helix_maxima}
\end{equation}
where \(\phi(0)=0\).

\begin{figure}[t]
\centering
\includegraphics[width=0.56\linewidth]{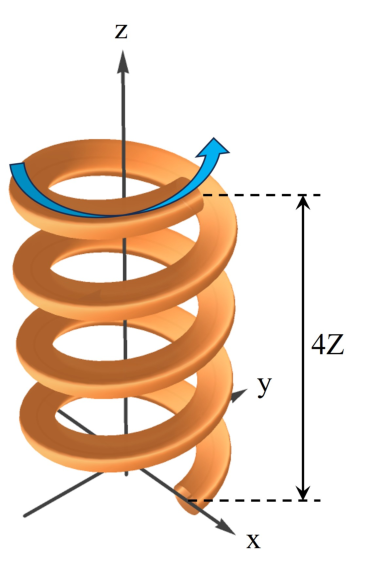}
\caption{A propagating Dirac electron along the \(+z\) direction with intrinsic spin up exhibits a helical conserved current at \(l=0\). The arrow indicates the circulation direction, and the label \(4Z\) marks four helix pitches. The helix represents a spatial current geometry rather than a particle trajectory. The central helical curve corresponds to the maximal azimuthal-current streamline at the characteristic radius \(\rho=\rho_*\), while the surrounding shaded tube provides a visual thickness around this streamline (rendered with radius \(a=0.2\,\rho_*\)) and does not represent the current magnitude or radial distribution. Reversing the electron spin yields the opposite helical handedness.}

\label{fig:helical_current}
\end{figure}

Figure~\ref{fig:helical_current} visualizes the intrinsic helical geometry of the conserved Dirac current at \(l=0\). This section establishes that a propagating Dirac electron exhibits an intrinsic, spin-resolved helical conserved current at \(l=0\), forming a spatial geometry that persists into the evanescent region.

\section{Helix Pitch and Local Chiral Coupling}

The existence of an intrinsic helical current structure in a propagating electron implies that a chiral environment can couple locally to the electron's spatial current geometry, without mediation by orbital motion. In this sense, chirality functions as a geometric \emph{selector} for the intrinsic helical-current modes already present in the Dirac current. Importantly, this selection is a helicity-dependent \emph{scattering/coupling} process that may involve spatial phase matching and mode conversion, whereas the details are presented in a follow-up work.

The geometric scale governing such coupling is set by the spatial pitch of the helical current structure. This pitch is determined by the azimuthal advance of a current streamline at fixed radius \(\rho\), which evolves linearly with the longitudinal coordinate \(z\) according to Eq.~\eqref{eq:helix_solution}. We define the Dirac spin--current helix pitch \(Z(\rho)\) as the longitudinal distance over which the azimuthal angle advances by \(2\pi\),
\begin{equation}
Z(\rho)\,
\frac{\zeta J_1(\zeta\rho)}{\rho\,k\,J_0(\zeta\rho)}
= 2\pi .
\end{equation}
Evaluated at the characteristic radius \(\rho_*\), this yields
\begin{equation}
Z(\rho_*) \simeq 2.28\,\mathrm{nm}.
\end{equation}

The helix pitch is a fraction of the longitudinal de~Broglie wavelength \(\lambda_z \simeq 7.76\,\mathrm{nm}\) and therefore constitutes an \emph{independent geometric length scale}, rather than a phase effect. Because its magnitude is comparable to characteristic molecular helical pitches, a chiral environment can couple directly to this intrinsic current geometry through local spatial matching.

The helix pitch \(Z\) depends only on basic physical parameters, including the electron kinetic energy and the confinement geometry, and therefore serves as a geometric control of the helical structure relevant for local chiral coupling.

\section{Discussion and Conclusion}

We have shown that a propagating electron with intrinsic spin carries a helical
conserved current even in the absence of orbital angular momentum. This helix is
neither a particle trajectory nor a phase artifact, but a geometric organization
of the Dirac wave in real space with direct electromagnetic significance. From
this perspective, chirality-induced spin selectivity~\cite{NaamanWaldeck2015,
Yang2021} emerges naturally as a geometric mode-selection process: a chiral
environment couples locally to the intrinsic helical-current modes already
present in a propagating Dirac electron, producing helicity-dependent scattering
outcomes.

The helix pitch \(Z\) sets the microscopic length scale governing such coupling.
For realistic parameters, \(Z\) can be substantially shorter than the
longitudinal de~Broglie wavelength, confirming that the helical current geometry
constitutes an independent spatial structure rather than a phase effect. The
dependence of \(Z\) on basic physical parameters therefore provides a direct
geometric handle for controlling chiral spin selectivity.

More broadly, the intrinsic helical current structure identified here shows that
spin--momentum correlations arise directly from Dirac wave structure, logically
prior to band formation or many-body effects. Because this structure resides in
the spatial flow of a conserved current rather than in an internal two-level
Hilbert space, it cannot be captured within reduced spin-only descriptions that
eliminate spatial degrees of freedom. Instead, the present results point to a
geometric current picture in which chiral environments act as spatial probes that
resolve the intrinsic helicity and pitch of the travelling electron. This local,
geometric perspective is consistent with recent work treating electron spin as a
spatially extended, physically addressable degree of freedom at the single-
electron level~\cite{GaoShen2025_JPhysComm}, and suggests broader relevance across
nanoscale and molecular systems.

\begin{acknowledgments}
The authors thank Prof.\ Wei Wang for helpful discussions on spin and chirality in biological and biophysical systems.
\end{acknowledgments}

% \bibliography{references}

\bibliography{Helix}

\end{document}